\documentclass[preprint]{revtex4}
\usepackage{color}
\usepackage{epsfig}
\usepackage{amsmath}
\usepackage{epstopdf} 
\usepackage{mathrsfs}
\usepackage{graphicx}
\usepackage{ulem}
\usepackage{inputenc}

\begin{document}

\title{Photon correlations in the collective emission of hybrid gold-(CdSe/CdS/CdZnS) nanocrystal supraparticles}

\author{V. Blondot$^{1}$, D. G\'erard$^{1}$, G. Quibeuf$^{1}$, C. Arnold$^{1}$, A. Delteil$^{1}$, A. Bogicevic$^{2}$, T. Pons$^{2}$, N. Lequeux$^{2}$, S. Buil$^{1}$, J-P. Hermier$^{1,}$}
\email{jean-pierre.hermier@uvsq.fr}
\affiliation{$^{1}$ Universit\'e Paris-Saclay, UVSQ, CNRS, GEMaC, 78000, Versailles, France.\\
$^{2}$ Laboratoire de Physique et d'\'Etude des Mat\'eriaux, ESPCI-Paris, PSL Research University, CNRS UMR 8213, Sorbonne Universit\'e, 10 rue Vauquelin, 75005 Paris, France}

\begin{abstract}
We investigate the photon statistics of the light emitted by single self-assembled hybrid gold-CdSe/CdS/CdZnS colloidal nanocrystal supraparticles through the detailed analysis of the intensity autocorrelation function $g^{(2)}(\tau)$. We first reveal that, despite the large number of nanocrystals involved in the supraparticle emission, antibunching can be observed. We then present a model based on non-coherent F\"orster energy transfer and Auger recombination that well captures photon antibunching. Finally, we demonstrate that some supraparticles exhibit a bunching effect at short time scales corresponding to coherent collective emission.
\end{abstract}

\maketitle

\section{Introduction}
Superradiance is a collective emission phenomenon predicted more than half a century ago by R. Dicke \cite{Dicke54}. Since the 70's, the concept of superradiance is at the origin of several multi-faceted works in astrophysics concerning the amplified diffusion of light \cite{Bekenstein73, Bekenstein98, Brito15}. In the field of optics, it has also attracted a great attention, motivating fundamental physics works \cite{Rehler71,Gross82} as well as studies about the principles of new laser devices \cite{Bonifacio89}.

Due to the inherent homogeneity of their properties, trapped atoms are a system of choice to experimentally implement the concept of superradiance \cite{Ortiz18}. Regarding condensed-matter emitters, the progress in material nanostructuration or the use of photonic structures offer the possibility to tune light matter interaction and to achieve superradiance for a wide range of nanoemitters such as molecules \cite{Luo19}, nanotubes \cite{Doria18} or diamond color centers \cite{Vass22}. 

In order to promote coherent collective emission, plasmonic nanocavities are particularly well suited. They allow to confine and tune the electromagnetic field at subwavelength scales. The plasmonic mode can play a crucial role, mediating and adjusting the coupling in the near field \cite{Pustovit09,Pustovit10,Shlesinger18}. This approach can also be extended to obtain phase locking between 2D arrays of spasers \cite{Dorofeenko13}.

In this paper, we show the collective emission of  hybrid plasmonic/colloidal quantum dot structures. They consist in self-assemblies of core-shell colloidal CdSe/CdS/CdZnS nanocrystals (NCs) encapsulated into a silica shell and a gold nanoresonator. The first section of the paper summarizes the chemical synthesis and the main properties of these golden supraparticles (GSPs) concerning their photoluminescence decay rate and  non coherent F\"orster energy transfer (FRET) between single NCs. By performing a detailed time-resolved analysis of the intensity autocorrelation function $g^{(2)}(\tau)$, we then show that photon antibunching is observed for several GSPs. This result is well modeled by a Monte-Carlo simulation taking into account FRET as well as Auger recombinations and a fraction of non emitting NCs. More interestingly, we evidence that the light emitted by some GSPs exhibits bunching at short delays, suggesting coherent collective emission.

\section{Gold-(CdSe/CdS/CdZnS) nanocrystal supraparticles: synthesis and basic optical properties}
\begin{figure*}[htp]
\centering
\includegraphics [width=15cm]{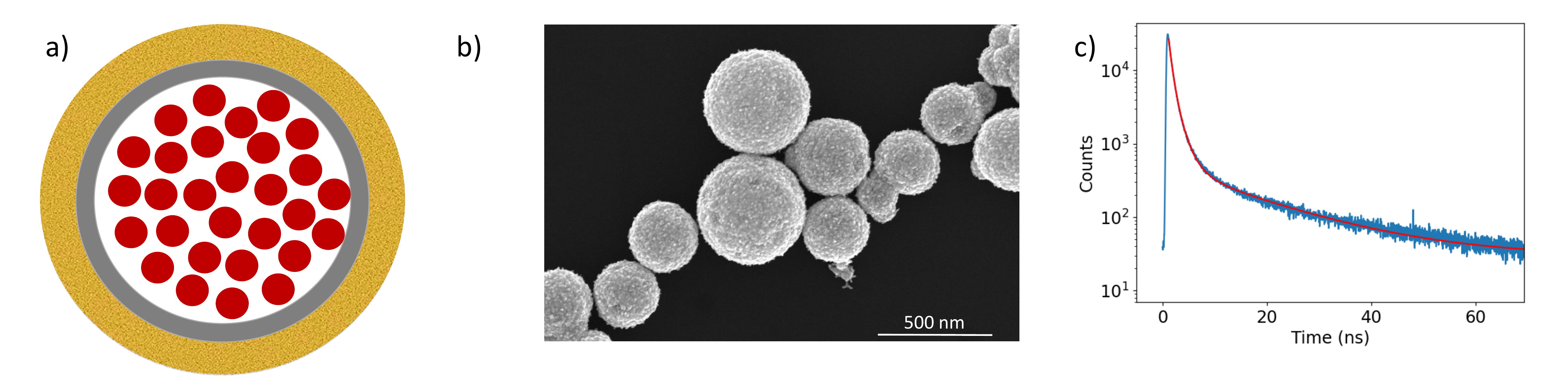}
\caption{Schematic representation of a single GSP (a).  Wide field images of GSPs (b). Typical PL decay rate of an individual GSP (c). It is fitted with a log-normal distribution with a central decay rate $\tau_{LN}$= 1 ns and a slow exponential component ($\tau$ = 17 ns) corresponding to the contribution of non radiative traps.}\label{figure_1}
\end{figure*}

A procedure with several steps was implemented to synthesize hybrid gold CdSe/CdS/CdZnS colloidal NC supraparticles as detailed in \cite{Bogicevic2022}. Briefly, each GSP includes from a few hundred to a few thousand NCs with a diameter of 7.7 $\pm$ 1 nm (fluorescence centered at 645 nm, FWHM = 30 nm). Following the approach presented in \cite{Blondot20}, the NC aggregate is first encapsulated in a silica shell with a thickness of 15 nm. It prevents quenching by the gold nanoshell of the emission of the NCs located at the surface of the aggregate \cite{Lakowicz2005}. After functionalization of the silica layer, gold seeds are then deposited according to the synthesis reported by Halas {\it et al.} \cite{Oldenburg1998}. The synthesis of continuous gold nanoshells with a thickness of 19 $\pm$ 5 nm is finally achieved (see Figure \ref{figure_1}).

The gold nanoshell first results in an increase of the PL decay rate through the well-known Purcell effect that was measured and modeled in detail \cite{Blondot23}. Depending on the GSP diameter, the Purcell factor ranges between 3 and 8, corresponding to PL lifetime between 0.7 ns and 1.9~ns (see Figure \ref{figure_1}.c). Close packing the NCs in a compact aggregate also results in the possibility of F\"orster resonant energy transfer (FRET) through the well-known process described by F\"orster. The small NCs (with the shortest fluorescence wavelength) act as donors for the larger ones (with a long fluorescence wavelength). With respect to aggregates without gold nanoshell, we showed that the gold nanoshell inhibits the contribution of FRET to the total decay rate of the smallest GSPs by a factor around 3 \cite{Blondot23}. In the next section, we investigate the opportunities opened by the reduction of this incoherent process in order to achieve coherent interactions between close NCs.

\section{Light emission of the GSPs - photon statistics}
\subsection{Photon counting and confocal microscopy setups}
\begin{figure*}[htp]
\centering
\includegraphics [width=15cm]{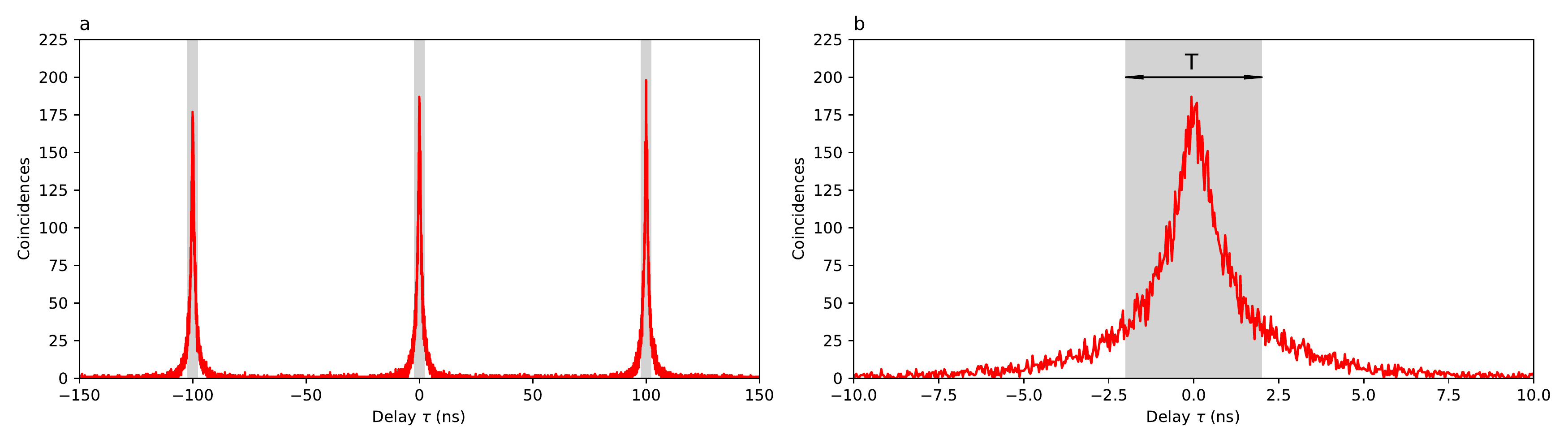}
\caption{(a) Schematic representation of the $g^{(2)}(\tau)$ time-integration method. For a given value $T$, we sum the number of coincidences recorded during a time window $T$ centered around the maximum of each peak \cite{Bradac17}. (b) Zoom of the central peak.}\label{figure_2}
\end{figure*}
40 µL of the GSP water solution is directly deposited on a glass coverslip where a TiO$_2$ grid with numbered cells of $65~\mu m \times 65~\mu m$ was previously prepared by photolithography. A set of single GSPs is selected by using an atomic force microscope (AFM) before optical measurements. Finally, we  characterize their form by scanning electron microscopy (SEM). Only individual GSPs with a shape that is close to a sphere are further considered.

The sample is positioned inside a confocal microscope (Attocube, Attodry 1100, objective numerical aperture = 0.82) operating at 4 K  and equipped with piezoelectric positioners that enable to excite a selected GSP by a focused laser beam provided by a pulsed laser diode (Picoquant LDH 520, wavelength = 520 nm, FWHM 160 ps). A dichroic mirror and a fluorescence filter separate the reflected laser light and the GSP emission, which is collected by a fiber and sent to a standard Hanbury Brown and Twiss detection setup. The signal of the two avalanche photodiodes (MPD, time resolution of 50 ps) is recorded by an acquisition card (Picoquant, PicoHarp 300, time bin of 64 ps) also synchronized to the pulsed laser diode. From the same data set, the decay of the luminescence and the histogram of the delays between photons can be plotted, the latter providing the intensity autocorrelation function $g^{(2)}(\tau)$.

A pioneer work reporting superradiant emission with quantum dots was based on the measurement of the enhancement of the PL decay rate \cite{Scheibner07}. However, in the case of most condensed-matter emitters, this approach can lack reliability due to the possible generation of non-radiative traps during the fabrication process. Moreover, if cooperative emission is only achieved for a small fraction of emitters, the signal may be hidden by the standard fluorescence of the remaining emitters. In contrast, as in many quantum optics experiments, characterizing the intensity autocorrelation function $g^{(2)}(\tau)$ appears as a very robust approach to demonstrate collective emission through the prediction of super-Poissonian and even superthermal statistics \cite{Temnov09, Jahnke16, Bradac17}. As in \cite{Bradac17}, we analyzed the data by using a time resolved approach consisting in plotting the variations of the time-integrated function $g^{(2)}(\tau)$ (see figure \ref{figure_2}). More precisely, the area of the peak around zero delay is integrated over a duration $T$ and normalized by the mean area of the lateral peaks, integrated over the same duration $T$. This quantity is noted $\tilde{g}^{(2)}_T(0)$ and enables to reveal non-Poissonian statistics at short time scale when $T$ decreases.

\subsection{Results}
\begin{figure*}[htp]
\centering
\includegraphics [width=15cm]{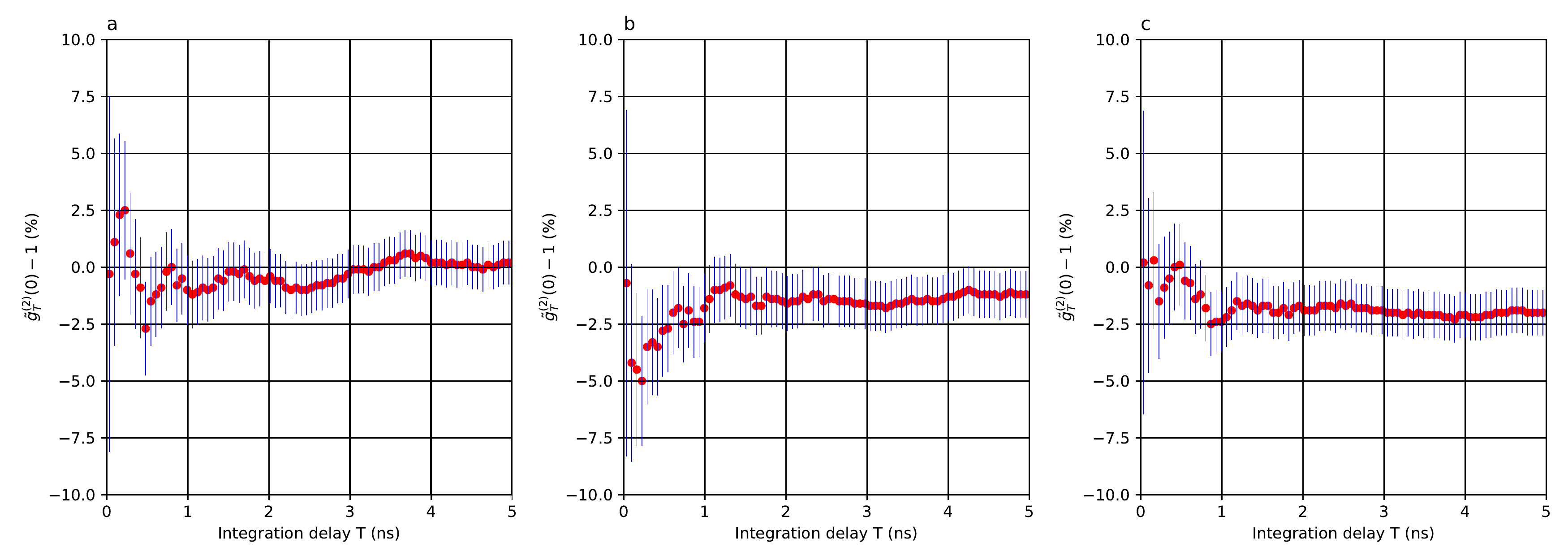}
\caption{Time-integrated function  $\tilde{g}^{(2)}_T(0)$ for 3 GSPs showing a Poissonian statistics (a) or a small amount of antibunching (b,c).  The error bars are calculated by taking into account	 the finite number of coincidences.}\label{figure_3}
\end{figure*}

Thanks to this analysis, different behaviors can be evidenced. The results for first three GSPs are presented in figure \ref{figure_3}. For some GSPs like (a), the photon statistics is Poissonian whatever the value of $T$. Some GSPs, as shown in (b) and (c), exhibit a small degree of antibunching that is not expected for non-interacting emitters. Indeed, if we consider $N$ dipoles emitting independently, it is well-known that $g^{(2)}(0)$ is equal to the value $(N-1)/N$, which can be used to develop postselection methods \cite{Mangum13}. However, the emission from clusters of colloidal NCs can exhibit stronger antibunching due to non-radiative energy transfer \cite{Whitcomb14}. Due to Auger recombinations, energy transfer between two excited NCs results in a first non-radiative decay channel. It can be compared to the Auger recombination of a biexcitonic state in a single NC, which is at the very origin of their single photon emission. Alternatively, blinking or photobleaching can create another non-radiative recombination channel since an excitation can be transferred from an "on" NC to an "off" one by FRET. 

In ref \cite{Whitcomb14}, the number of emitters is low ($\leq 4$) so that the authors could derivate a rate equation model to calculate the amount of antibunching. In the case of GSPs, such approach is not possible and we therefore performed Monte-Carlo simulations. We consider an ensemble of $N$ NCs arranged in a cubic lattice with periodic boundary conditions. The emission energy of each NC is different and is drawn randomly in a set of equally distributed values. When excited, a NC can transfer its energy by FRET to one of the 6 closest NCs if its fluorescence wavelength is lower than the considered adjacent NC. The corresponding rate $k_{FRET}$ is set to the same value as the radiative one (typical for such self-assembled NCs structures \cite{Blondot20,Blondot23}). In the case of a 6 nm CdS shell thickness, Auger recombinations at 4 K are partially inhibited due to electron delocalization into the shell and the radiative quantum efficiency is about 50 \% \cite{Canneson14}. Since the NCs used for the present study exhibit a thinner shell (1.6 nm), we consider that biexcitonic radiative recombinations do not occur. Even if blinking is nearly suppressed for such kind of emitters at 4 K \cite{Canneson14}, we reported previously that a fraction of the NCs are damaged during the aggregate synthesis \cite{Blondot20}. This is equivalent to the fraction $1-F$ of NCs in the "off" state taken into account in \cite{Whitcomb14}. More recently, an in-depth analysis of the first-order coherence of the light emitted by GPSs allowed us to numerically estimate the fraction $F$ of bright NCs in such a mesoscopic ensemble \cite{Delteil22}. It ranges between 10 \% to values exceeding 90 \%. 
\begin{figure*}[htp]
\centering
\includegraphics [width=15cm]{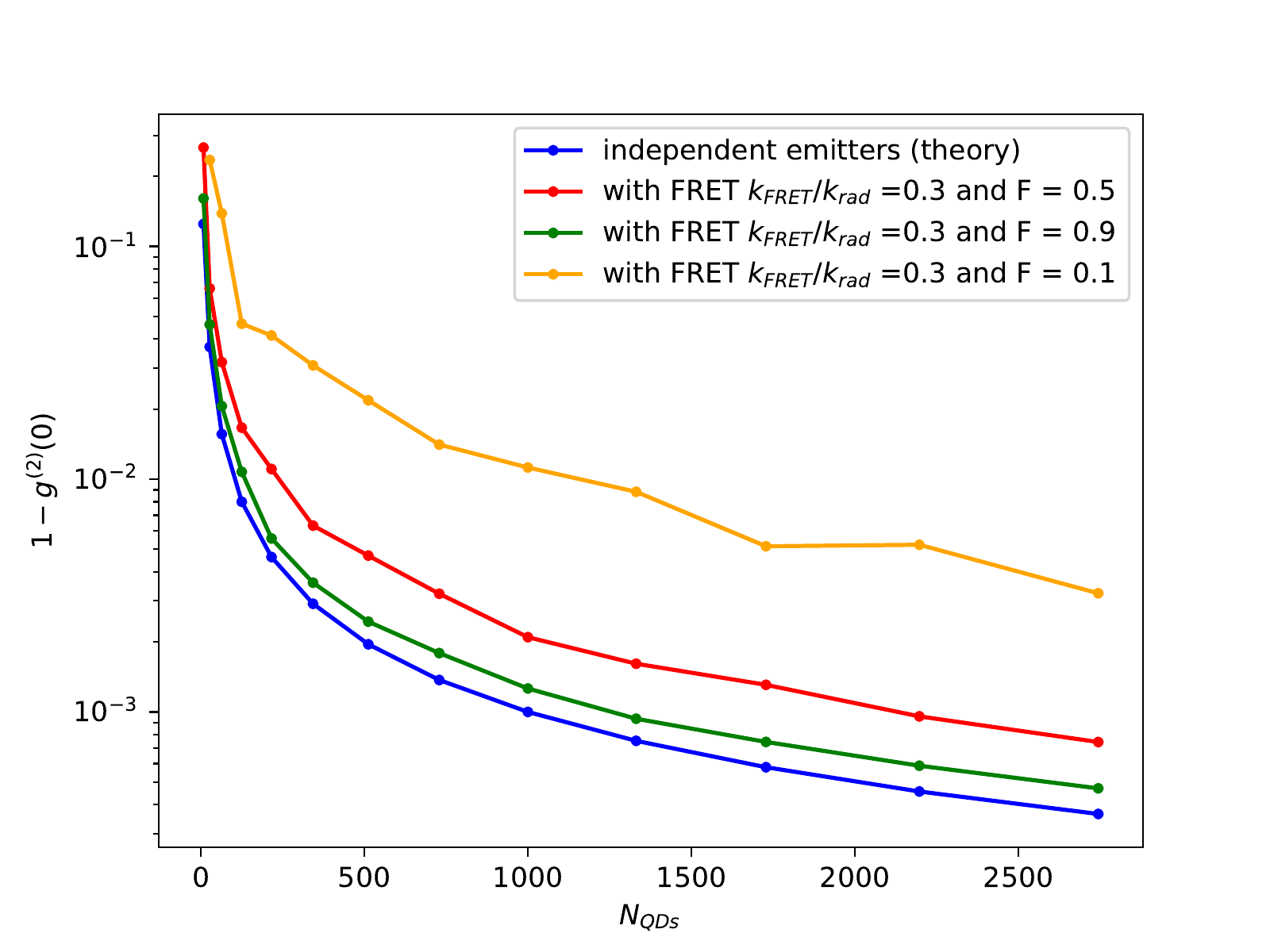}
\caption{Monte-Carlo simulations of $1-g^{(2)}(0)$ for a cubic crystal of $N$ NCs that can exchange energy by FRET with a rate $k_{FRET}$. The fraction of emitting NCs is $F$. $k_{rad}$ is the radiative decay rate. The probability to excite each NC per pulse is 0.3.}\label{figure_4}
\end{figure*}

Figure \ref{figure_4} shows the variations of $g^{(2)}(0)$ for various parameter values as a function of the number of NCs. The model predicts an excess amount of antibunching with respect to the $(N-1)/N$ value corresponding to $N$ independent emitters. As in \cite{Whitcomb14}, it increases with $F$ and it reaches values that are experimentally observed in Figure \ref{figure_3}. For some GSPs, the absence of antibunching is likely to be related to a strong reduction of FRET processes \cite{Blondot23}, large values of $F$ and/or bunching coming from collective emission. 

\begin{figure*}[htp]
\centering
\includegraphics [width=12cm]{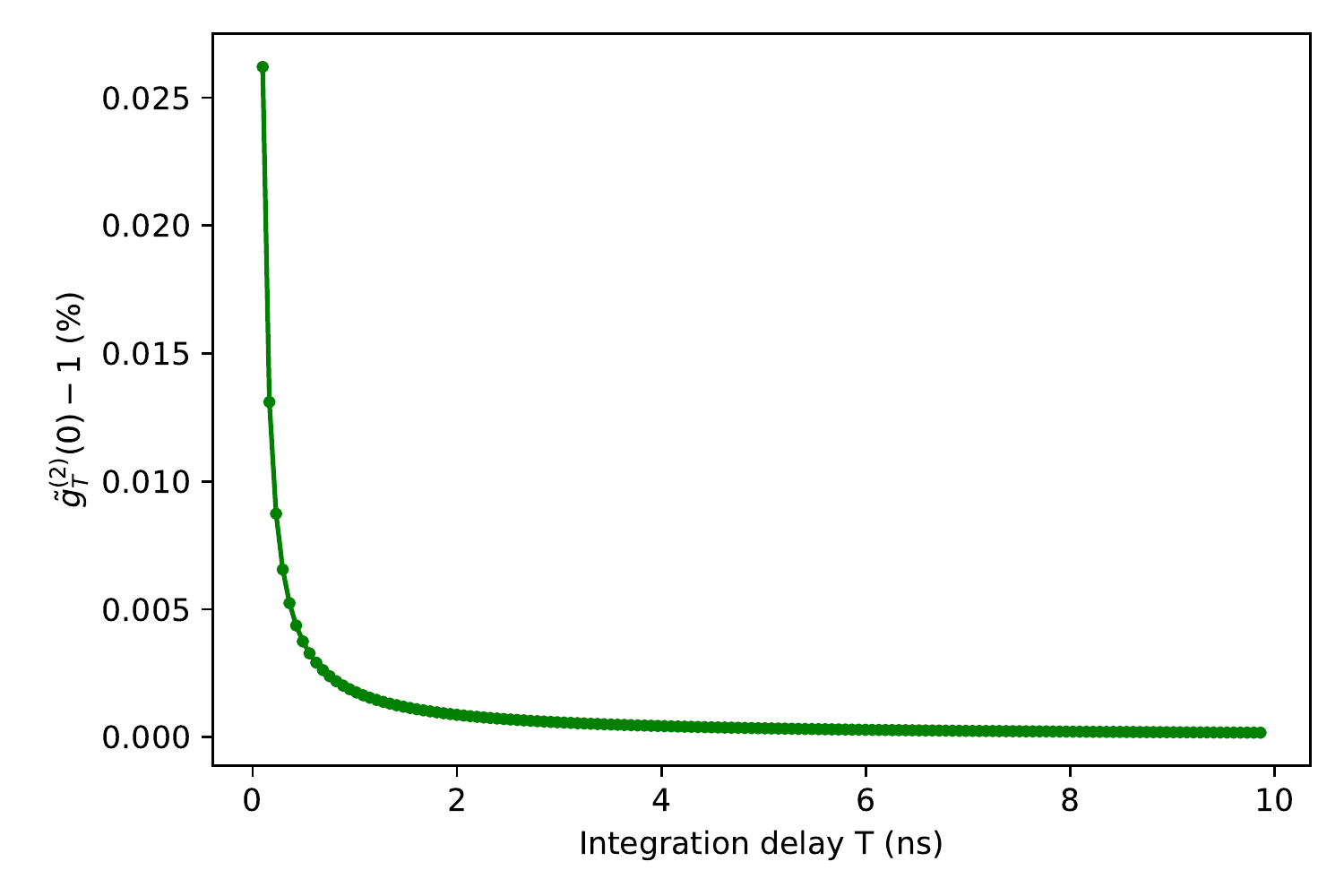}
\caption{Calculated time-integrated function  $\tilde{g}^{(2)}_T(0)$ for a thermal source with a linewidth of 10 THz (corresponding to the typical emission linewidth of a GSP).}\label{figure_6}
\end{figure*}

Before investigating this latter effect, we first evaluate the photon bunching we would measure from a standard thermal emission characterized by a value $g^{(2)}(0)=2$. Taking into account the time coherence ($\sim$ 100 fs) deduced from the linewidth of emission spectrum ($\sim$ 15 nm \cite{Blondot23}) as well as the time bin in our experiment (64 ps), we calculated the autocorrelation function $\tilde{g}^{(2)}_T(0)$ (see Figure \ref{figure_6}). The amount of bunching $\tilde{g}^{(2)}_T(0)-1$ would not exceed a negligible value of 0.03 \%.

\begin{figure*}[htp]
\centering
\includegraphics [width=15cm]{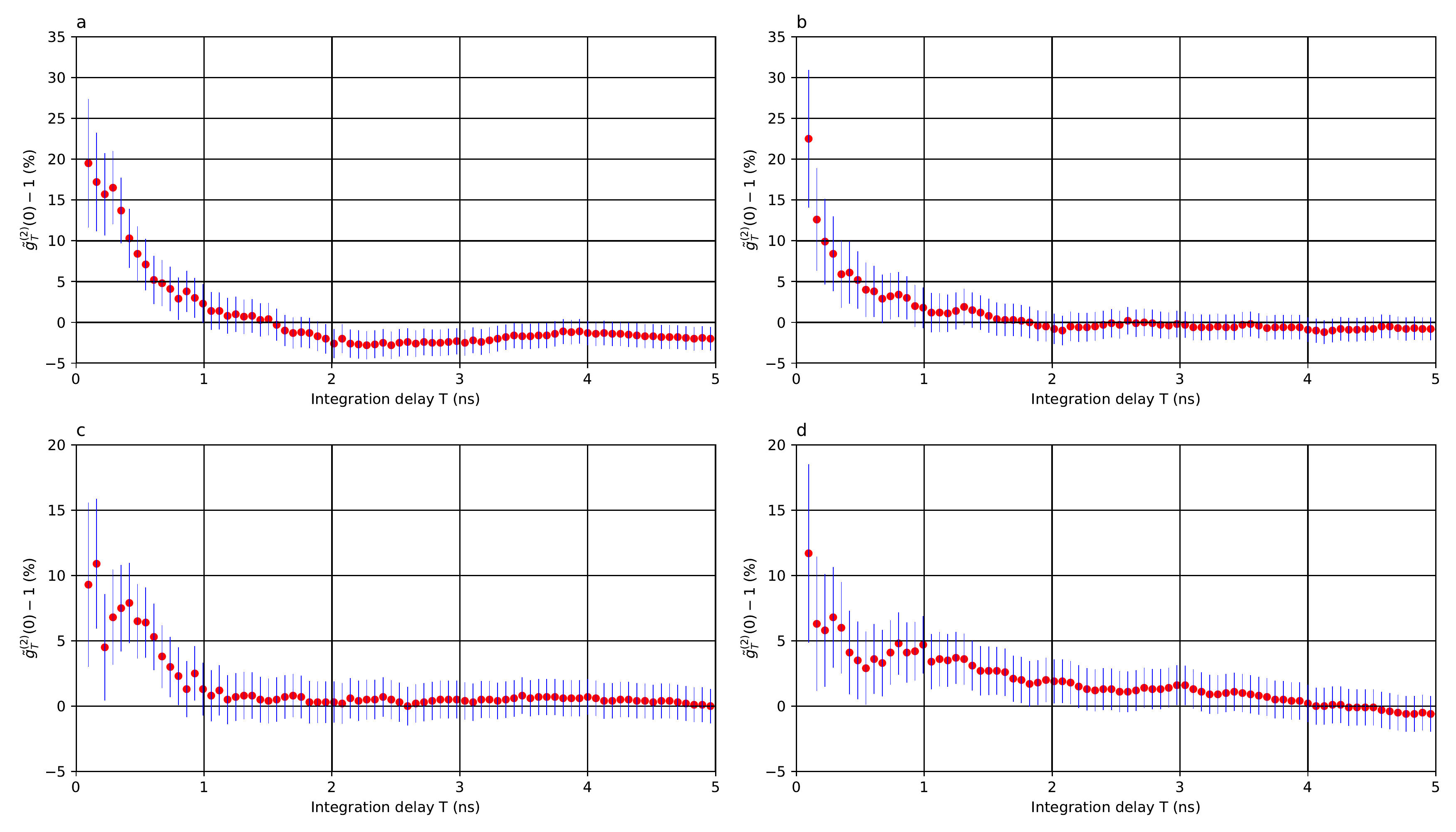}
\caption{Time-integrated function  $\tilde{g}^{(2)}_T(0)$ for 4 GSPs showing superthermal bunching.}\label{figure_5}
\end{figure*}

In contrast with previously shown, up to 20 \% bunching is observed for some GSPs at short values of $T$ (figure \ref{figure_5}). These results were only obtained using an excitation power 10 times lower than previously, in order to reduce as much as possible processes such as Auger recombinations and promote coherent interactions between NCs. The probability to excite one NC in the aggregate is then lower than 10 \% (this value is deduced from the evaluation of the setup collection efficiency, which is about 0.5 \%). 

When considering the ideal case of $N$ independent and identical dipoles coupled to a single cavity mode, the cooperativity $C$, which quantifies the emission into the cavity mode with respect to the other modes, is $N$ times higher than in a single emitter-in-cavity case \cite{Bonifacio76}. This property can also be seen as an enhancement by a factor $\sqrt{N}$ of the effective coupling rate between the emitters and the cavity mode. As a result, collective emission occurs and superradiance  is achieved with an emission decay rate $N$ times higher with respect to the single emitter case \cite{Dicke54}. Let us consider an ensemble of emitters with a fluorescence decay rate $\gamma$ and coupled to an environment inducing dephasing with a rate $\gamma^\ast$. Qualitatively, when coupled to a cavity with a Purcell factor $F_P$, collective effects can be achieved if $\gamma^\ast$ is smaller than $N F_P \gamma$ so that collective emission occurs before the processes at the origin of the decoherence of the dipoles act \cite{Meiser10}.

Concerning GSPs, the dephasing rate for a single GSP is typically $\gamma^\ast=$ 10 THz at 4 K \cite{Blondot23,Delteil22} while $F_p \gamma=$ 1 GHz (the mean lifetime is 1 ns, see previous section), leading to the prediction that collective effects can indeed take place if about $10^4$ NCs are involved. This value can be reached for standard GSPs: taking into account the structure of the GSPs, the NC diameter, and the volume fraction (66 \%) occupied by the NCs \cite{Zaccone22}, one finds that the number of NCs ranges between 5000 (diameter $\sim$ 220 nm) and 75000 (diameter $\sim$ 400 nm). While significant, the amount of bunching (20 \%) decreases with $T$, suggesting that only a fraction of NCs are involved in collective emission processes. A large subset of NCs experience non-collective (and therefore slower) light emission showing no bunching. As a result, $\tilde{g}^{(2)}_T(0)$ approaches 0 or negative values as $T$ increases. The absence of bunching for large values of $T$ also demonstrates the relevance of the time-resolved approach. Not only does it overcome the limitations of standard methods such as PL decay rate measurements, but it allows us to discriminate between bunching at short time scales and Poissonian or antibunching for longer time scales.

\section{Conclusion}
This work focuses on the analysis of the statistical properties of the light emitted by individual hybrid gold/colloidal NCs structures (GSPs). Using a specific time-resolved method to analyze the intensity autocorrelation function $g^{(2)}(\tau)$, we demonstrate that the amount of bunching at short delays can reach 20 \% for a single GSP, showing collective emission. At long time scales, antibunching is sometimes observed. It fundamentally comes from energy transfer between adjacent NCs and non radiative recombinations involved in such NCs aggregates. The realization of hybrid gold-colloidal nanostructures based on nanoscale emitters with much higher oscillator strength, such as nanoplatelets \cite{Ithurria11}, could open the possibility of achieving collective emission at room temperature.

\section*{Conflicts of interest}
There are no conflicts to declare.

\section*{Acknowledgements}
This work was funded by the Agence Nationale de la Recherche in the framework of the GYN project (Grant No. ANR-17-CE24-0046).


\begin{thebibliography}{99}
\bibitem{Dicke54} R. H. Dicke, "Coherence in Spontaneous Radiation Processes," Phys. Rev. {\bf 93}, 99 (1954)
\bibitem{Bekenstein73} J. Bekenstein, "Extraction of energy and charge from a black hole," Phys.Rev. D {\bf 7}, 949 (1973)
\bibitem{Bekenstein98} J. D. Bekenstein and M. Schiffer, "The Many faces of superradiance," Phys.Rev. D {\bf 58}, 064014 (1998)
\bibitem{Brito15} R. Brito, V. Cardoso, and P. Pani, "Superradiance, New Frontiers in Black Hole Physics", 2nd ed., Springer (2020)
\bibitem{Rehler71} N. E. Rehler and J. H. Eberly, "Superradiance," Phys. Rev. A {\bf 3}, 1735 (1971)
\bibitem{Gross82} M.Gross and S.Haroche, "Superradiance: An essay on the theory of collective spontaneous emission," Phys. Rep {\bf 93}, 301 (1982)
\bibitem{Bonifacio89} R. Bonifacio, B. W. J. Mc Neil, and P. Pierini, "Superradiance in the high-gain free-electron laser," Phys. Rev. A {\bf 40}, 4467 (1989)
\bibitem{Ortiz18} L. Ortiz-Guti\'errez, L. F. Mu\~{n}oz-Mart\'inez, D. F. Barros, J. E. O. Morales, R. S. N. Moreira, N. D. Alves, A. F. G. Tieco, P. L. Saldanha, and D. Felinto "Experimental Fock-State Superradiance," Phys. Rev. Lett. {\bf 120}, 083603 (2018)
\bibitem{Luo19} Y. Luo, G. Chen, Y. Zhang, L. Zhang, Y. Yu, F. Kong, X. Tian, Y. Zhang, C. Shan, Y. Luo, J. Yang, V. Sandoghdar, Z. Dong, and J. G. Hou, "Electrically Driven Single-Photon Superradiance from Molecular Chains in a Plasmonic Nanocavity," Phys. Rev. Lett. {\bf 122}, 233901 (2019)
\bibitem{Doria18} S. Doria, T. S. Sinclair, N. D. Klein, D. I. G. Bennett, C. Chuang, F. S. Freyria, C. P. Steiner, P. Foggi, K. A. Nelson, J. Cao, A. Aspuru-Guzik, S. Lloyd, J. R. Caram, and M. G. Bawendi, "Photochemical Control of Exciton Superradiance in Light-Harvesting Nanotubes," ACS Nano {\bf 12}, 4556 (2018)
\bibitem{Vass22} D. Vass, A. Szenes, B. B\'anhelyi, and M. Csete, "Plasmonically Enhanced Superradiance of Broken-Symmetry Diamond Color Center Arrays Inside Core-Shell Nanoresonators," Nanomaterials {\bf 12}, 352 (2022) 
\bibitem{Pustovit09} V. N. Pustovit and T. V. Shahbazyan, "Cooperative emission of light by an ensemble of dipoles near a metal nanoparticle: The plasmonic Dicke effect," Phys. Rev. Lett. {\bf 102}, 077401 (2009)
\bibitem{Pustovit10} V. N. Pustovit and T. V. Shahbazyan, "Plasmon-mediated superradiance near metal nanostructures," Phys. Rev. B {\bf 82}, 075429 (2010)
\bibitem{Shlesinger18} I. Shlesinger, P. Senellart, L. Lanco, and J.-J. Greffet, "Tunable bandwidth and nonlinearities in an atom-photon interface with subradiant states," Phys. Rev. A {\bf 98}, 013813 (2018)
\bibitem{Dorofeenko13} A. V. Dorofeenko, A. A. Zyablovsky, A. P. Vinogradov, E. S. Andrianov, A. A. Pukhov, and A. A. Lisyansky, "Steady state superradiance of a 2D-spaser array," Opt. Expr. {\bf 21}, 14539 (2013)
\bibitem{Bogicevic2022} A. Bogicevic, X. Xu, S. Buil, C. Arnold, J.-P. Hermier, T. Pons, and N. Lequeux,  "Synthesis and characterization of colloidal quantum dot superparticles - plasmonic gold nanoshell hybrid nanostructures," DOI:10.26434/chemrxiv-2022-1m9q5
\bibitem{Blondot20} V. Blondot, A. Bogicevic, A. Coste, C. Arnold, S. Buil, X. Qu\'elin, T. Pons, N. Lequeux, J.-P. Hermier, "Fluorescence properties of self assembled colloidal supraparticles from CdSe/CdS/ZnS nanocrystals," New J. Phys. {\bf 22}, 113026 (2020)
\bibitem{Lakowicz2005} J. R. Lakowicz,  "Radiative decay engineering 5: metal-enhanced fluorescence and plasmon emission," Anal. Biochem. {\bf 337}, 171 (2005)
\bibitem{Oldenburg1998} S. J. Oldenburg, R. D. Averitt, S. L. Westcott, and N. J. Halas, "Nanoengineering of optical resonances," Chem. Phys. Lett. {\bf 288}, 243 (1998)
\bibitem{Blondot23} V. Blondot, C. Arnold, A. Delteil, D. G\'{e}rard, A. Bogicevic, T. Pons, N. Lequeux, J.-P. Hugonin,  J.-J. Greffet, S. Buil, J-P. Hermier, "Fluorescence decay enhancement and FRET inhibition in self-assembled hybrid gold CdSe/CdS/CdZnS colloidal nanocrystals supraparticles," Opt. Expr. {\bf 31}, 4454 (2023)
\bibitem{Scheibner07} M. Scheibner, T. Schmidt, L. Worschech, A. Forchel, G. Bacher, T. Passow, and D. Hommel, "Superradiance of quantum dots," Nat. Phys. {\bf 3}, 106 (2007)
\bibitem{Temnov09} V. V. Temnov and U. Woggon, "Photon statistics in the cooperative spontaneous emission," Opt. Expr. {\bf 17} 5774-5782 (2009)
\bibitem{Jahnke16} F. Jahnke, C. Gies1, M. A\ss mann, M. Bayer, H.A.M. Leymann, A. Foerster, J. Wiersig, C. Schneider, M. Kamp and S. H\"ofling, "Giant photon bunching, superradiant pulse emission and excitation trapping in quantum-dot nanolasers," Nat. Comm. {\bf 7}, 11540 (2016)
\bibitem{Bradac17} C. Bradac, M. T. Johnsson, M. van Breugel, B. Q. Baragiola, R. Martin, M. L. Juan, G. K. Brennen, and T. Volz, "Room-temperature spontaneous superradiance from single diamond nanocrystals," Nat. Comm. {\bf 8} 1205 (2017)
\bibitem{Mangum13} B. D. Mangum, Y. Ghosh, J. A. Hollingsworth, and H. Htoon, "Disentangling the effects of clustering and multi-exciton emission in second-order photon correlation experiments," Opt. Expr. {\bf 21}, 7419 (2013)
\bibitem{Canneson14} D. Canneson, L. Biadala, S. Buil, X. Quélin, C. Javaux, B. Dubertret, J.-P. Hermier, "Blinking suppression and biexcitonic emission in thick-shell CdSe/CdS nanocrystals at cryogenic temperature," Phys. Rev. B. {\bf 89}, 035303 (2014)
\bibitem{Whitcomb14} K. J. Whitcomb, J. Q. Geisenhoff, D. P. Ryan, M. P. Gelfand, and A. Van Orden, "Photon Antibunching in Small Clusters of CdSe/ZnS Core/Shell Quantum Dots," J. Phys. Chem. B {\bf 119}, 9020 (2015)
\bibitem{Delteil22} A. Delteil, V. Blondot, S. Buil, and J.-P. Hermier, "First-order coherence of light emission from inhomogeneously broadened mesoscopic ensembles," Phys. Rev. B {\bf 106}, 115302 (2022)
\bibitem{Bonifacio76} R. Bonifacio, and L. Lugiato, "Cooperative effects and bistability for resonance fluorescence," Opt. Commun. {\bf 19}, 172 (1976)
\bibitem{Meiser10} D. Meiser and M. J. Holland, "Steady-state superradiance with alkaline-earth-metal atoms," Phys. Rev. A {\bf 81}, 033847 (2010)
\bibitem{Zaccone22} A. Zaccone, "Explicit Analytical Solution for Random Close Packing in d=2 and d=3," Phys. Rev. Lett. {\bf 128}, 028002 (2022)
\bibitem{Ithurria11} S. Ithurria, M. D. Tessier, B. Mahler, R. P. S. M. Lobo, B. Dubertret, and Al. L. Efros, "Colloidal nanoplatelets with two-dimensional electronic structure," Nat. Mat. {\bf 10}, 936 (2011)

\end{thebibliography}
\end{document}